\documentclass[english,a4paper]{article}
\usepackage{dcolumn}
\usepackage{bm}
\usepackage{graphicx}
\usepackage{amsmath}
\usepackage{amsfonts}
\usepackage{amssymb}
\usepackage{amsthm}
\usepackage{amstext}
\usepackage{amsbsy}
\usepackage{amsopn}
\usepackage{amscd}
\usepackage{amsxtra}

\newtheorem{theorem}{Theorem}

\newtheorem{proposition}{Proposition}

\def\openone{\leavevmode\hbox{\small1\kern-3.3pt\normalsize1}}

\usepackage[usenames,dvipsnames]{color}

\newcommand{\XX}{{\bf X}}
\newcommand{\PP}{{\bf P}}

\begin{document}
\title{Chattering Phenomenon in Quantum Optimal Control}
\author{R. Robin\footnote{Laboratoire Jacques-Louis Lions, CNRS, Inria, Sorbonne Universit\'e, Universit\'e de Paris, France, remi.robin@inria.fr}, U. Boscain\footnote{Laboratoire Jacques-Louis Lions, CNRS, Inria, Sorbonne Universit\'e, Universit\'e de Paris, France, ugo.boscain@sorbonne-universite.fr}, M. Sigalotti\footnote{Laboratoire Jacques-Louis Lions, CNRS, Inria, Sorbonne Universit\'e, Universit\'e de Paris, France, mario.sigalotti@inria.fr}, D. Sugny\footnote{Laboratoire Interdisciplinaire Carnot de
Bourgogne (ICB), UMR 6303 CNRS-Universit\'e Bourgogne-Franche Comt\'e, 9 Av. A.
Savary, BP 47 870, F-21078 Dijon Cedex, France, dominique.sugny@u-bourgogne.fr}}

\maketitle

\begin{abstract}
We present a quantum optimal control problem which exhibits a chattering phenomenon. This is the first instance of such a process in quantum control.
Using the Pontryagin Maximum Principle and a general procedure due to V. F. Borisov and M. I. Zelikin,
we characterize the local optimal synthesis, which is then globalized by a suitable numerical algorithm.
We illustrate the importance of detecting chattering phenomena because of their impact on the efficiency of
numerical optimization procedures.
\end{abstract}

\section{Introduction}
\begin{figure}
\begin{center}
\includegraphics[scale=0.85]{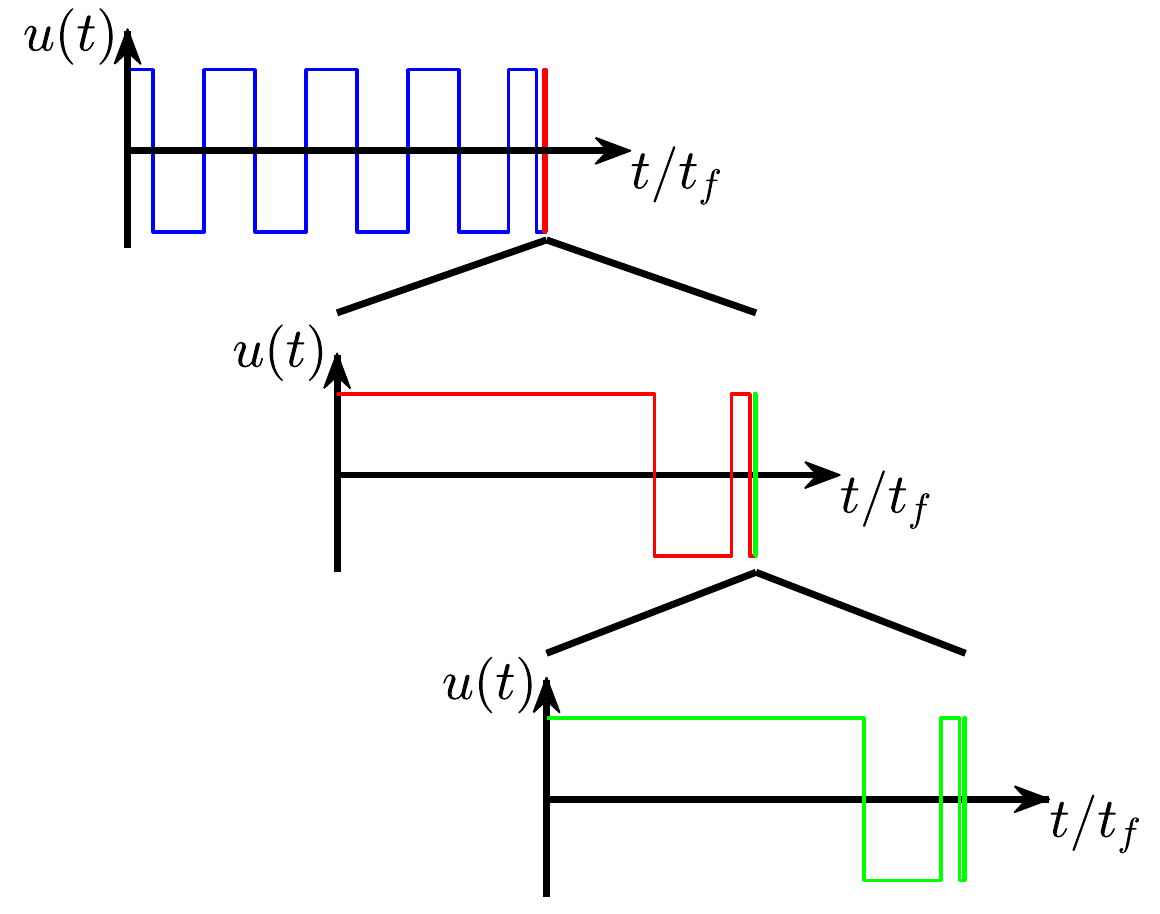}
\end{center}
\caption{Schematic description of the time scale invariance of the optimal control in the chattering process. The optimal solution of the quantum control problem described in this paper is plotted in the top panel. Near the final time $t_f$, the control switches infinitely many times with an asymptotic invariant structure by time dilation as represented on the two lower panels (successive zooms around $t=t_f$ given by the different plot colors).}\label{fig0}
\end{figure}

Consider the experiment in which a ball bounces up and down on the ground. We assume that the impact with the ground is inelastic and that the ball is only subject to the gravity. In the ideal case in which the ball changes its speed instantaneously at each bounce, an infinite number of bounces is performed in the finite time of the process.  Chattering thus refers to an observable (here the speed) having very fast 
discontinuities, which lead in the mathematical limit to an infinite number of jumps in a finite-time
interval~\cite{zelikinbook}. This type of process can also be found in quantum physics.  Examples are the quantum Zeno effect and dynamical decoupling in which
a repeated observation of the system and a periodic sequence of instantaneous pulses prevent, respectively, its time evolution~\cite{misra77} or its coupling with the environment~\cite{DD}.
The possibility of chattering was also established in Optimal Control Theory (OCT). OCT  was founded in the sixties by Pontryagin and his co-workers~\cite{pontryaginbook}, with a rigorous framework to design control protocols for driving a dynamical system from a given initial state into a desired target
state, while minimizing energy or other resources. Chattering was found in this field by A.T.~Fuller in a planar system~\cite{fuller,borisov,schaettler-book}. It consists of a control that switches infinitely many times over a finite time period~\cite{zelikinbook}. This observation runs counter to common
experience for which control is viewed as a piecewise continuous function, while for chattering, the control lies in a larger class of functions~\cite{boscain2021}. In Fuller's example, the number of switchings accumulates with a geometric progression at the final control time. The control law has a time scale invariance near the final time as schematically represented in Fig.~\ref{fig0}. At first Fuller's example was considered a curiosity, but this type of phenomenon is very widespread in optimal control, as rigorously shown few years later by I. Kupka~\cite{kupka1990}. While optimal
trajectories exhibiting  chattering  have been found in relevant examples from medicine~\cite{schattler2012} to classical~\cite{zelikinbook} and space dynamics~\cite{zhu2016}, this phenomenon has, to the best of our knowledge, not yet been studied in quantum control~\cite{cat,brifreview,roadmap,koch2019,alessandrobook}. The existence, the role and the ubiquity of this process in quantum systems remain an open question. Such control schemes are interesting from a fundamental point of view even if they turn out to not be feasible in experiments. They may also have a rather severe impact on the numerical search of optimal solutions~\cite{zhu2016,brysonbook}. It is therefore important to understand why chattering is occurring and how it can be avoided~\cite{caponigro2018}. We propose in this paper to describe this phenomenon in quantum control by studying a simple but fundamental quantum system. We introduce on this key example a systematic procedure to design the optimal control protocol. The impact on the efficiency of optimization procedures is also described.
\section{The model system}
Let us consider the control of a three-level quantum system described by a pure state belonging to a Hilbert space spanned by the states $|1\rangle$, $|2\rangle$, and $|3\rangle$. As in a standard STIRAP process~\cite{RMPstirap}, in a suitable rotating frame and in the rotating wave approximation,
the dynamics of the system are controlled by two pulses $\Delta$ and $u$ that couple, respectively, states $|1\rangle$ and $|2\rangle$ and states $|2\rangle$ and $|3\rangle$. We assume that $\Delta$ is a constant and that the only control parameter is $u=u(t)$. The resulting dynamics are
\begin{equation}
  \label{eq:dyn}
  \dot{\XX}=(\Delta\Omega_3+u(t)\Omega_1)\XX,
\end{equation}
where $\XX$ is a vector of real coordinates $(x_1,x_2,x_3)$ with the condition $x_1^2+x_2^2+x_3^2=1$
(see~\ref{sec3level}). 
The two skew-symmetric matrices $\Omega_1$ and $\Omega_3$ are given by
$$
\Omega_1=\left(
\begin{array}{ccc}
0 & 0 & 0\\
0 & 0 & -1\\
0 & 1 & 0
\end{array}\right),\quad
\Omega_3=\left(\begin{array}{ccc}
0 & -1 & 0\\
1 & 0 & 0\\
0 & 0 & 0
\end{array} \right),
$$
generating respectively the rotations along the $x_1$- and $x_3$-directions. In the following, the  control cannot exceed a certain physical bound, that is,  $|u(t)|\leq u_0$ for some $u_0>0$. A time rescaling results in the multiplication of  the two parameters $\Delta$ and $u_0$ of the problem by a positive scalar, which leads to the normalization $u_0=1$.

Starting from any state $\XX_0$ on the unit sphere, the goal of the control is to steer the system to  the state $(0,0,1)$, still denoted by $|3\rangle$, while minimizing the population transfer to the state $|1\rangle$. This control protocol is interesting in practice if, e.g., the state $|1\rangle$ is the only state of the system subject to an unwanted relaxation process. Notice that this latter is not modeled by Eq.~(\ref{eq:dyn}). The control protocol can be formalized as an optimal control problem by introducing the cost functional $\mathcal{C}=\int_0^{t_f}x_1^2(t)dt$ to be minimized, where $t_f$ is the control time, which is not fixed.  The existence of a solution in finite time is not obvious and it is a consequence of the general theory developed in~\cite{zelikinbook}, as explained in~\ref{seczelikin}. 
We give below an argument showing that, once its existence has been established, such a control has a chattering behaviour.
\section{Application of the Pontryagin Maximum Principle}\label{sec:PMP}
The existence of chattering can be proved from the Pontryagin Maximum Principle (PMP), which gives a first-order necessary condition for optimality~\cite{boscain2021}. It can be stated by introducing the Pontryagin Hamiltonian
$$
H_P
= \PP\cdot(\Delta\Omega_3+u\Omega_1)\XX+p^0 x_1^2,
$$
 where $\XX$ is a point on the unit sphere, $\PP$  the adjoint state of coordinates $(p_1,p_2,p_3)$, and $p^0$  a constant equal either to 0 or to $-\frac12$. If $\XX(t)$ is an optimal trajectory with corresponding control $u(t)$, then $\XX(t)$ is \emph{extremal}, namely, there exist $\PP(t)$ and $p^0$ such that
 $\dot \XX=\frac{\partial H_P}{\partial \PP}$,
 $\dot \PP=-\frac{\partial H_P}{\partial \XX}$, and $H_P(\XX(t),\PP(t),u(t),p^0)=\max_{v\in [-1,1]}H_P(\XX(t),\PP(t),v,p^0)=0$.
Moreover,
  $(\tilde \PP(t),p^0)\neq(0,0)$ for every $t$, where $\tilde \PP(t)=\PP(t)-(\PP(t)\cdot \XX(t))\XX(t)$.
  If $p^0=0$ (resp. $p^0=-\frac12$) the extremal is called \emph{abnormal} (resp. \emph{normal}).
   The condition $(\tilde \PP(t),p^0)\neq(0,0)$ (instead of the 
  weaker but more classical  one $(\PP(t),p^0)\neq(0,0)$)
  reflects the fact that 
   we already know that $\XX(t)$ belong to the two-dimensional unit sphere. Actually, $(\XX(t),\tilde \PP(t))$ represents an element of the cotangent bundle of the sphere. 

The maximization condition of the PMP is solved with the switching function $\Phi(t)=\PP(t)\cdot \Omega_1\XX(t)$.
 If $\Phi(t)$ is different from zero  then $H_P$ is maximal when the control, called \emph{bang}, is a constant control of maximum amplitude, $u(t)= \textrm{sign}[\Phi(t)]$.
  If $\Phi(t)\ne 0$ on a time-interval $(a,b)$, we say that the restriction of the extremal to $(a,b)$ is a \emph{bang arc}.
  When
   $\Phi$ is constantly equal to zero on $(a,b)$ then the restriction of the extremal to $(a,b)$ is said to be a \emph{singular arc}.
%
Singular arcs can be characterized as follows. A simple computation shows that $\dot\Phi=\Delta\, \PP\cdot \Omega_2\XX$,  where $\Omega_2$ is the generator of the rotations along the $x_2$- axis.
If both $\Phi$ and $\dot\Phi$ vanish at time $t$ then either $\tilde\PP(t)=0$ or $\Omega_1\XX(t)$ and $\Omega_2\XX(t)$ are linearly dependent, that is $\XX(t)$ lies on the equator $x_3=0$. If $\tilde \PP$ is equal to zero on a time interval  then $p^0=-\frac12$ and, moreover, $0=H_P=-\frac{x_1^2}{2}$, leading to $x_1=0$ on the same interval. Since $\dot x_1=-\Delta x_2$, it follows that $\XX$ stays at $|3\rangle$.




We now claim that it is enough to consider normal extremals, i.e., that any optimal trajectory corresponding to an abnormal extremal also corresponds to a normal one.
Indeed, if an optimal trajectory reaches $|3\rangle$ in a finite time $T\le t_f$ then, taking $\hat t_f>t_f$, the extension ${\bf Y}$ of $\XX$ on $[0,\hat t_f]$ that  stays in $|3\rangle$ on the interval $[t_f,\hat t_f]$ is
optimal as well. Moreover, ${\bf Y}$ corresponds to a vanishing control for all times in $[T,\hat t_f]$. Let us denote by ${\bf Q}$ the
adjoint state corresponding to ${\bf Y}$, whose existence is guaranteed by the PMP.
Then the extremal ${\bf Y}$ is singular on $[T,\hat t_f]$ and  $\tilde {\bf Q}|_{[T,\hat t_f]}\equiv 0$, where $\tilde {\bf Q}$ is defined in analogy to $\tilde \PP$. Such an extremal cannot be abnormal (otherwise $(\tilde {\bf Q}(\hat t_f),p^0)$ would be zero).
We have shown that for any optimal solution reaching $|3\rangle$ in time $T$
it can be assumed
without loss of generality that 
 the extremal is normal and $\tilde \PP|_{[T,t_f]}\equiv 0$ (this corresponds to
replacing $\PP$ by ${\bf Q}|_{[0,t_f]}$ and denoting the latter by $\PP$).

The last step consists in showing that an optimal trajectory cannot be the finite concatenation of bang arcs
in $[0,T]$, where we recall that $T$ is the time at which $\XX$ reaches $|3\rangle$. This fact can be deduced by computing the so-called order of the singular trajectory and using general properties of optimal control theory (see, e.g.,~\cite{schattler2012,Lewis,Lev}).
In our specific case we can exhibit an elementary and explicit proof.
By contradiction, assume that $u$ is constantly equal to $+1$ or to $-1$ in an interval $[T-\varepsilon,T]$ for some $\varepsilon>0$. Then $\Phi$ is smooth on $[T-\varepsilon,T]$ and an explicit computation  based on the dynamics
\begin{equation}\label{eq:gamiltonian}
\dot p_1=-\Delta\, p_2+x_1,\  \dot p_2=\Delta\, p_1-u\,p_3,\  \dot p_3=u\, p_2
 \end{equation}
 and on the relation $\tilde\PP(T)=0$
shows that $\Phi(T)=\dot \Phi(T)= \Phi^{(2)}(T)=\Phi^{(3)}(T)=0$ and  $\Phi^{(4)}(T)=-u\Delta^2$ (see~\ref{secswitching}).  
Then $\Phi(t)$ has opposite sign with respect to $u$ in a small left neighborhood of $T$, contradicting the fact that  $u(t)=\textrm{sign}[\Phi(t)]$. Hence the only option to reach $|3\rangle$ is  a chattering process in which the control switches infinitely many times in a finite time-interval.
Notice that, as a byproduct, we obtain that there are no optimal abnormal extremals reaching the target. Indeed, a simple computation shows that the control corresponding to an abnormal extremal reaching the target at time $T$ is necessarily bang on an interval $[T-\varepsilon,T]$ for $\varepsilon>0$ small enough (since $\Phi(T)$ must be zero and $\dot \Phi(T)$ cannot be zero at the same time).

Having established the chattering behaviour, the next goal is to find the position of the switching points, which form the switching curve. This is not an easy task and an exact analytic expression cannot be derived. Notice however that close to the state $|3\rangle$ the system can be described by the two coordinates $x_1$ and $x_2$ with the dynamics
\begin{equation}
\dot x_1=-\Delta\, x_2,~\dot x_2=\Delta\, x_1-u\sqrt{1-x_1^2-x_2^2}.\label{ooo1}
\end{equation}
Taking only the dominant terms, the system can be locally approximated as
\begin{equation}
\dot x_1=-\Delta\, x_2,~\dot x_2= -u,\label{ooo2}
 \end{equation}
 which is different from the standard linear approximation. In control literature, system~(\ref{ooo2}) is referred to as the \emph{nilpotent approximation} of (\ref{ooo1}) \cite{ABB}.
The dynamics of (\ref{ooo2}) together with the cost $\int_0^{t_f}x_1^2(t)\,dt$ ($t_f$ free) and the origin $x_1=x_2=0$ as target state,  yield  the  classical Fuller  problem,  up to the change of coordinates $x_1\to -x_1/\Delta$, $x_2\to x_2$.
The chattering trajectories of the Fuller model can be described exactly~\cite{schaettler-book} as recalled in~\ref{secfuller}. 
We then deduce the following properties for 
system~(\ref{ooo2}). 
The optimal solution is bang-bang with an infinite number of switchings near the origin. The switching curve is defined by $x_1=\textrm{sign}[x_2]\xi \Delta x_2^2$,
where $\xi=\sqrt{\frac{\sqrt{33}-1}{24}}\simeq 0.44623 $. The switching times $t_k$ are given by a geometric progression of the form $\frac{T-t_k}{T-t_{k-1}}=\frac{1}{\alpha}$ with $\alpha=\sqrt{\frac{1+2\xi}{1-2\xi}}\simeq 4.1301$.

The relation between the local optimal syntheses of two systems differing only by high-order terms (such as (\ref{ooo1}) and (\ref{ooo2})),
when one of the two syntheses exhibits chattering, is a subtle question.
To this purpose
we  apply  the results of~\cite{zelikinbook}, which permit to conclude that a system has a Fuller-like optimal synthesis provided it differs from the Fuller model by terms that are small while applying suitable non-isotropic dilations. On the basis of this study described in~\ref{seczelikin}, 
we state the following result.
\begin{proposition}%
For every point $\XX_0$ sufficiently close to $|3\rangle$,
any optimal solution of (\ref{eq:dyn}) connecting $\XX_0$ to $|3\rangle$ corresponds to a control
having infinitely many discontinuities accumulating at the first time $T$
at which $|3\rangle$ is reached (and possibly staying at $|3\rangle$ for larger times).
The optimal synthesis is characterized by a switching curve $\Gamma$ passing through $|3\rangle$, whose expression in coordinates $(x_1,x_2)$ is
of the form
\begin{eqnarray*}
  x_1=\left\{\begin{array}{ll} \lambda_1(x_2) x_2^2\ \mbox{if }x_2>0,\\
\lambda_0(x_2) x_2^2\ \mbox{if }x_2<0,
\end{array}\right.
\end{eqnarray*}
 where $\lambda_0$ and $\lambda_1$ are $\mathcal{C}^1$ function satisfying $\lambda_0(0)=
    -\lambda_1(0)=\Delta \xi$.
  In the coordinates $(x_1,x_2)$ and locally near $(0,0)$
   the optimal control is $-1$ below the curve $\Gamma$ and $+1$ above it.
\end{proposition}
\section{Numerical results}
The optimal synthesis can be computed numerically starting from that of the Fuller model. When we are sufficiently close to $|3\rangle$,
we approximate the switching curve of the quantum system by that of its  approximation~(\ref{ooo2}). We consider a point of the latter curve of coordinates $(\xi \Delta x_{20}^2,x_{20})$, with $x_{20}>0$ (notice that the same method could be used for $x_{20}<0$). The third component 
is obtained from $x_{30}=\sqrt{1-x_{10}^2-x_{20}^2}$ and the adjoint state from the condition $\PP\cdot \Omega_1\XX=0$ together with
 $H_P=0$~\cite{boscain2021}. The dynamics of the PMP allows to propagate $\XX$ and $\PP$ backward in time, uniquely up to an irrelevant component of $\PP$ along $\XX$. Starting from an initial point with $x_{20}>0$, one integrates the equations taking $u=-1$.  When the corresponding switching function vanishes, the control switches from $-1$ to $+1$. Then one goes on by integrating the equations with $u=+1$ up to the next switching time and so on. The stability of this approach is justified in~\ref{sec:gradient}. An optimal control law can be obtained for each value of $x_{20}$. Even if the result applied above to characterize  the optimal synthesis (see Theorem 1 in~\ref{seczelikin}) 
 can be used only for initial points $\XX_0$ close to $|3\rangle$, numerical simulations show that optimal trajectories have the same structure everywhere in the north hemisphere.

\begin{figure}
\begin{center}
\includegraphics[scale=0.75]{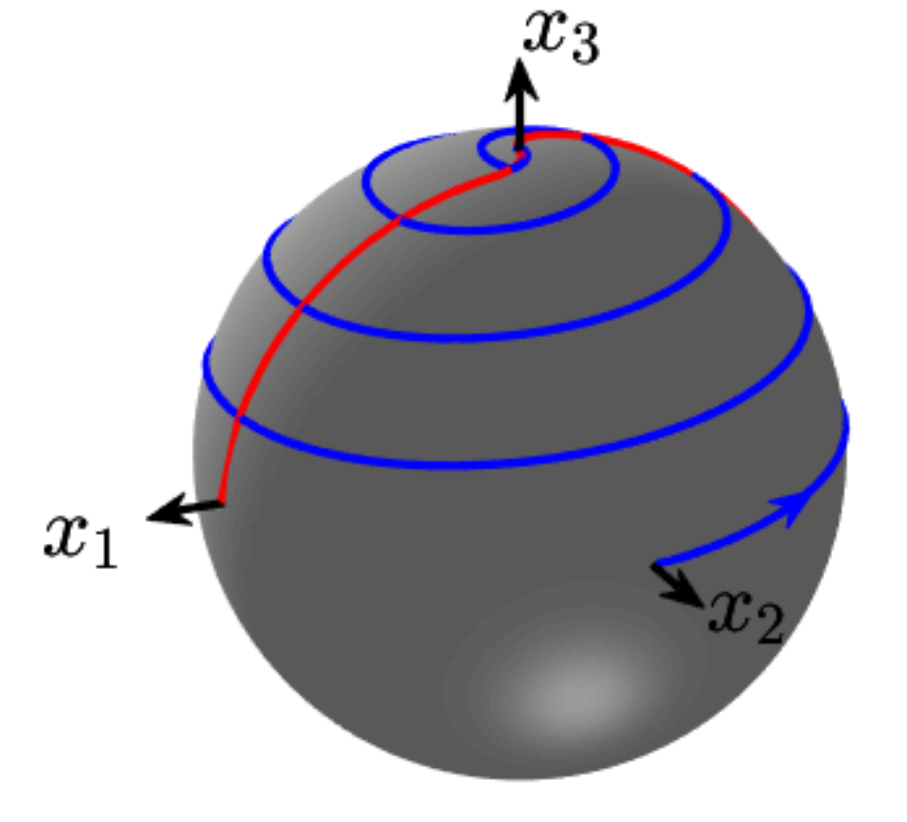}
\end{center}
\caption{Optimal trajectory from $\XX(0)=(0,1,0)$ to $\XX(t_f)=|3\rangle=
(0,0,1)$ on the sphere $x_1^2+x_2^2+x_3^2=1$. The switching curve is plotted in red. The parameter $\Delta$ is set to~10.\label{fig1}}
\end{figure}

\begin{figure}
\begin{center}
\includegraphics[scale=0.65]{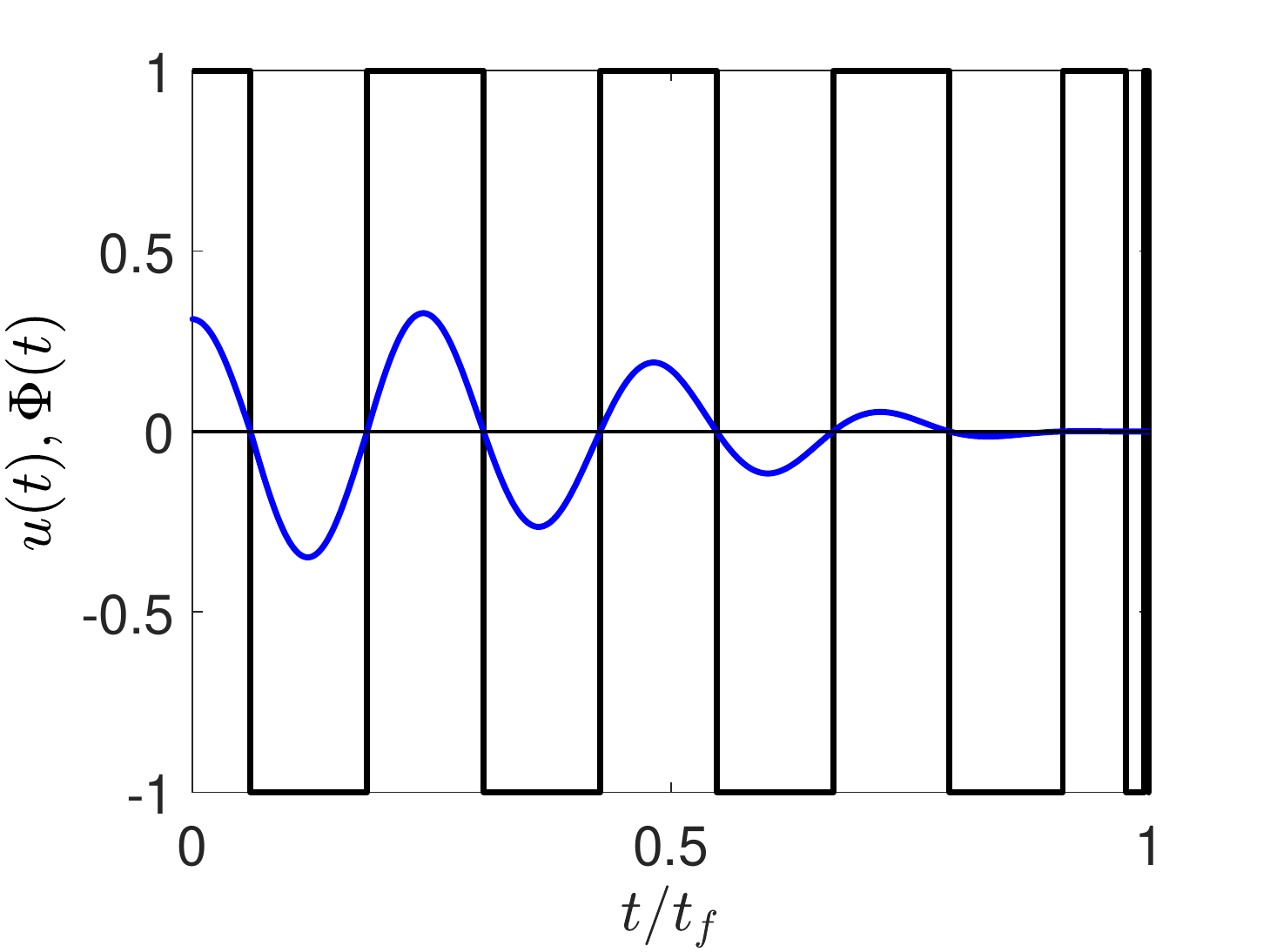}
\end{center}
\caption{Time evolution of the control $u(t)$ (bold black line) and of the switching function $\Phi(t)$ (blue line) for the optimal trajectory of Fig.~\ref{fig1}. The control switches from $\pm 1$ to $\mp 1$ when the switching function changes sign.\label{fig2}}
\end{figure}

The trajectory starting from a given initial state (say $(0,1,0)$) 
can then be determined by a Newton algorithm to estimate the right parameter $x_{20}$
 from which the backward propagation arrives at the  initial state. The parameter $x_{20}$ is not unique because the forward optimal trajectory has infinitely many switching points near the target. In practice, the choice of $x_{20}$  is dictated by the required precision on the final state. For $\Delta=10$ and a precision of $10^{-3}$, numerical simulations lead to $x_{20}=6.9\times 10^{-4}$. Figure~\ref{fig1} depicts the optimal trajectory on the unit sphere. The corresponding control $u(t)$ and switching function $\Phi(t)$ are displayed in Fig.~\ref{fig2}.

\begin{figure}
\begin{center}
\includegraphics[scale=0.75]{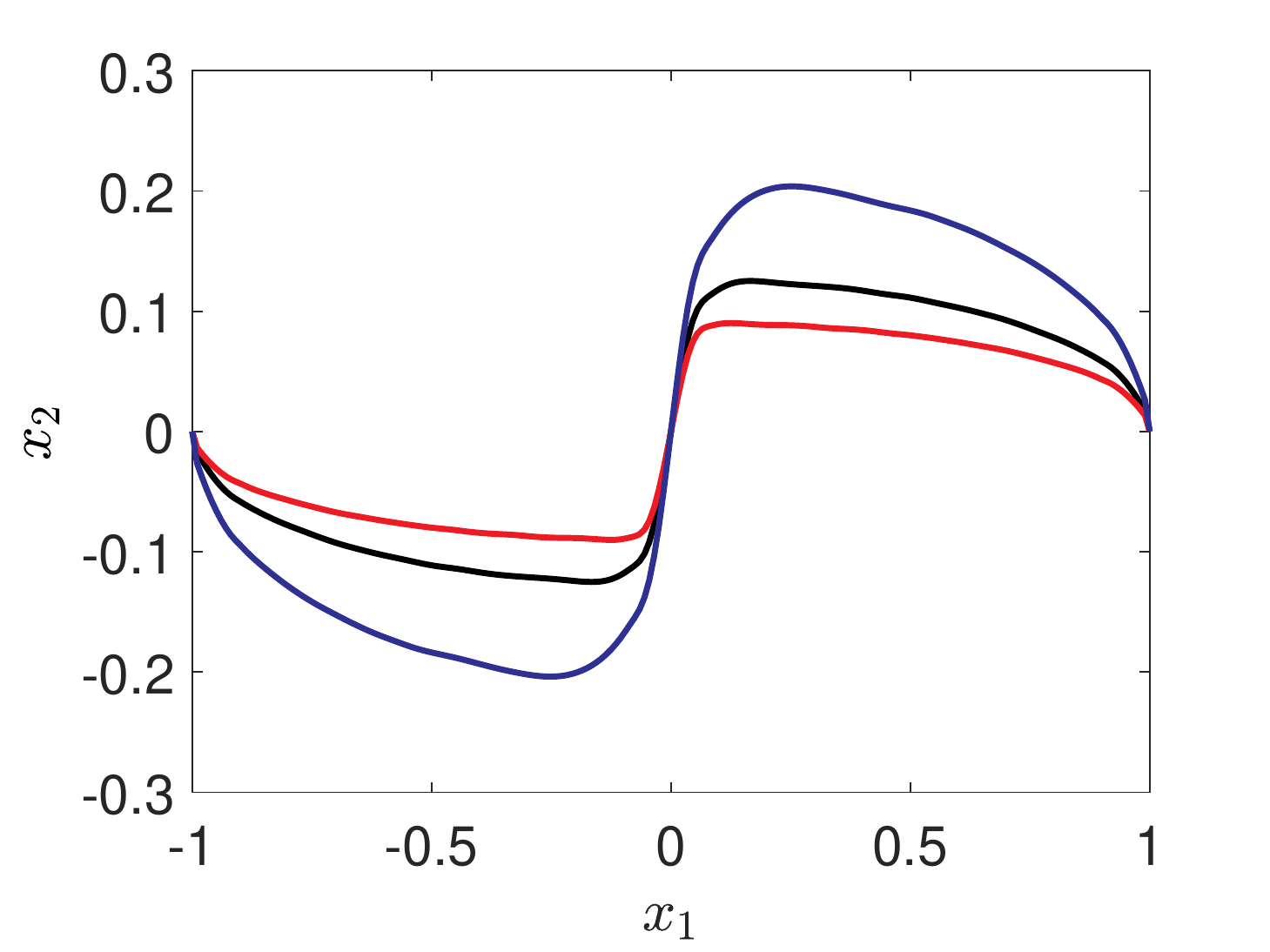}
\end{center}
\caption{Switching curves corresponding to three different
values of $\Delta$. The parameter is set respectively to 6 (blue), 10 (black), and 14 (red). \label{fig:switchingcurves}}
\end{figure}

The switching curves are reconstructed  numerically by  varying
the small parameter $x_{20}$ and collecting all the switching points of the corresponding trajectories obtained by integrating the extremal flow backwards in time. The result is represented in Fig.~\ref{fig1} and \ref{fig:switchingcurves}. 
We observe in Fig.~\ref{fig:switchingcurves} that the points $(\pm 1,0,0)$ belong to the switching curves of the quantum system, independently of the value of $\Delta$.
It can be shown that if this curve goes out of the north hemisphere then it passes through the points $(\pm 1,0,0)$ (see~\cite[Section 4.3]{BoscainChitourSIAM}, \cite{boscain-book} and \ref{secswitching}). 

\begin{figure}
\begin{center}
\includegraphics[scale=1]{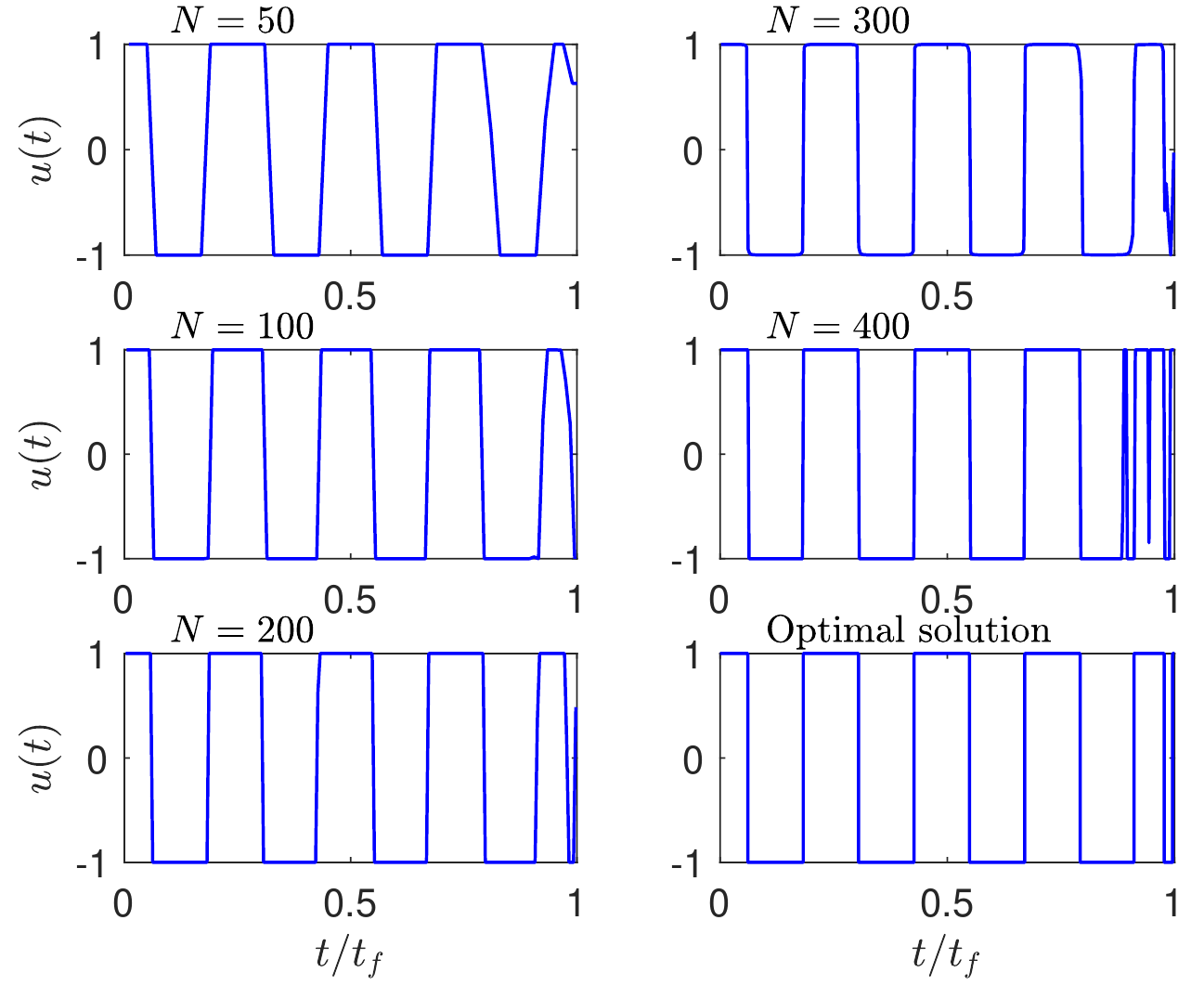}
\end{center}
\caption{Comparison between the
optimal solution of the PMP (bottom right panel) and a numerical control designed by a direct approach for different time steps $N$.\label{fig5}}
\end{figure}

The geometric analysis gives the optimal control protocol with a very high numerical precision. Due to their complexity, quantum control problems are usually solved by numerical optimization algorithms in which the control is assumed to be a piecewise constant function~\cite{cat}. An intriguing question is therefore to study to what extent such solutions can approximate the chattering phenomenon. The numerical simulations presented below use a direct optimal method with the software BOCOP~\cite{bocop} with a fixed control time computed using our analysis.
In Fig.~\ref{fig5}, we observe that the chattering process can only be reproduced approximately by numerical optimization. Additional results can be found in~\ref{sec:gradient}. 
The fineness of the time discretization corresponding here to $t_f/N$, where $N$ is the number of time steps, is a key factor to improve the protocol efficiency and to reproduce the control shape. However, for $N=400$, we observe an erratic structure of the control. Without a precise understanding of the optimal control strategy, this switching accumulation could be misinterpreted as numerical instabilities or artifacts, while it is due to the very structure of the optimal control. Reasonable efficiencies of the control protocols are achieved for quite small values of $N$. Such sub-optimal strategies could be used to bypass the problem due to chattering in the numerical optimizations.
\section{Conclusion}
In summary, the analysis of a simple system has
shown that chattering can appear in quantum optimal control. Such a process may play a key role in numerical optimization algorithms, as illustrated in the example by the rapid oscillations that occur in the numerical solutions.
In accordance with the existing results on the ubiquity of such phenomena in OCT~\cite{kupka1990}, we expect chattering to appear in many further examples, especially for higher-dimensional problems. Actually, one can give sufficient conditions on the relations between the commutators of the uncontrolled and controlled Hamiltonians
to guarantee the existence of chattering solutions (see \cite{kupka1990} and \cite[Chapter 4]{zelikinbook}). Such conditions render the chattering phenomenon more and more frequent as the dimension grows, and our example shows that no obstacle to their appearance comes from the quantum structure of the control problem. It should be noticed, however, that one cannot deduce from the general conditions in high dimensions
 the optimality of the chattering trajectories, unlike the two-dimensional system considered in this paper.\\

\noindent\textbf{Acknowledgment}\\
This research has been partially supported by the ANR project ``QUACO'' ANR-17-CE40-0007-01 and received funding from the European Union Horizon 2020 research and innovation program under Marie-Sklodowska-Curie Grant No. 765267 (QUSCO). This research has been supported by the ANR-DFG project CoRoMo and the ANR project QuCoBEC ANR-22-CE47-0008-02. \\

\noindent\textbf{Data availability statement}\\
The data that support the findings of this study are available upon reasonable request from the authors.

\appendix


\section{Derivation of the model system}\label{sec3level}
\begin{figure}
\begin{center}
\includegraphics[scale=0.75]{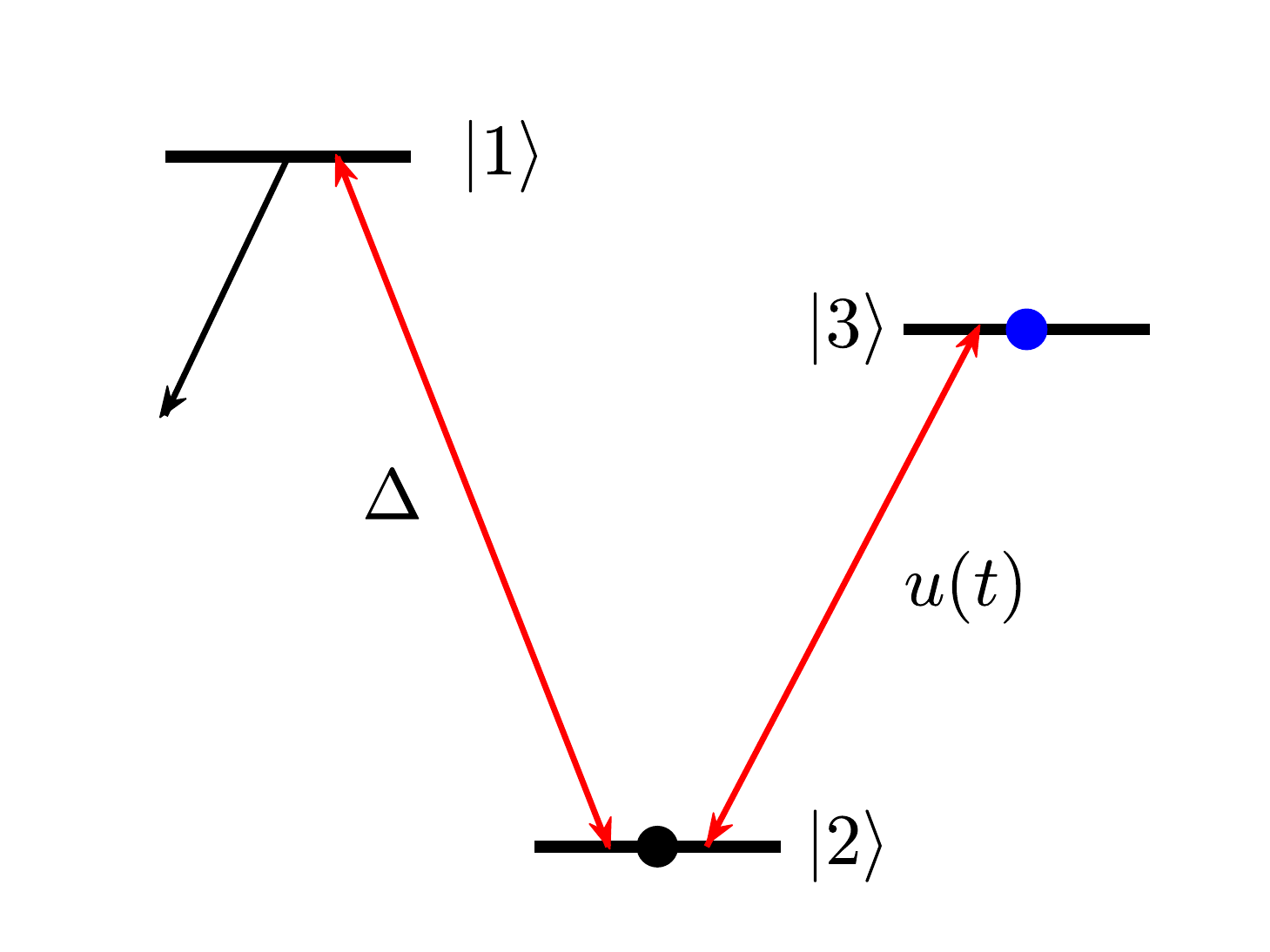}
\end{center}
\caption{Schematic representation of the quantum system with the coupling $\Delta$ and $u(t)$ between the states $|1\rangle$ and $|2\rangle$ and $|2\rangle$ and $|3\rangle$ (red arrows). The black and blue dots indicate respectively the initial and the target states. The black arrow represents the relaxation process.\label{figS0}}
\end{figure}

We show in this section how to derive the system used in this study. We consider a three-level quantum system in a $\Lambda$ configuration whose dynamics are governed by the Schr\"odinger equation. The
system is described by a pure state $|\psi(t)\rangle$ which belongs to a three-dimensional Hilbert space spanned by the basis $\{|1\rangle,|2\rangle,|3\rangle\}$. The system is subject to a pump and a Stokes pulses coupling, respectively, states $|1\rangle$ and $|2\rangle$ and states $|2\rangle$ and $|3\rangle$.
They  are assumed to be on-resonance with the corresponding frequency transitions.
 There is no direct coupling between levels $|1\rangle$ and $|3\rangle$. A schematic representation of the control problem is given in Fig.~\ref{figS0}.

The time evolution of the system is given by the Schr\"odinger equation
$$
i\frac{\partial}{\partial t}|\psi(t)\rangle = H(t)|\psi(t)\rangle,
$$
where units such that $\hbar =1$ are used. In the interaction representation and in the rotating-wave approximation, the Hamiltonian of the system can be written as
$$
H(t)=\left(\begin{array}{ccc}
0 & \Delta & 0\\
\Delta & 0 & u(t)\\
0 & u(t) & 0
\end{array}\right),
$$
where $\Delta$ and $u(t)$ are half the
 Rabi frequencies 
 of the two pulses. We denote by $c_1$, $c_2$, and $c_3$ the complex coefficients of the state $|\psi(t)\rangle$, and we introduce the real coefficients $x_1,\dots,x_6$ defined by $c_1=x_1+ix_4$, $c_2=x_5-ix_2$, $c_3=x_3+ix_6$.
  Since $H(t)$ is Hermitian, we have that $\sum_{i=1}^6x_i^2$ is constant. Up to normalization, $\sum_{i=1}^6x_i^2=1$ in accordance to the description of a pure state.
Straightforward computations from the Schr\"odinger equation
show that the variables $x_1$, $x_2$, and $x_3$ are decoupled from $x_4$, $x_5$, and $x_6$.
For our purposes it is sufficient to study the dynamics of the first set of variables, which turn out to be
\begin{eqnarray*}
& &\dot{x}_1=-\Delta x_2, \\
& & \dot{x}_2=\Delta x_1-u(t)x_3, \\
& & \dot{x}_3=u(t)x_2.
\end{eqnarray*}
In the control problem, we consider an initial state of the dynamics such that $x_4=x_5=x_6=0$. We deduce that $x_1^2+x_2^2+x_3^2=1$ at any time $t$.
\section{The Fuller model}\label{secfuller}
This paragraph briefly describes the main results that can be established for the classical Fuller model. The interested reader will find the proofs of the different statements in textbooks of mathematical control theory~\cite{zelikinbook,schaettler-book}.

The Fuller model is an  optimal control problem in $\mathbb{R}^2$. The dynamics of the state
 $(x,y)$ are governed by the differential equations
\begin{eqnarray}\label{eqfuller}
& & \dot{x}=y, \\
& &\dot{y}=u,\nonumber
\end{eqnarray}
where the control $u=u(t)$ satisfies the constraint $u(t)\in [-1,1]$. Starting from the state $(x_0,y_0)$, the goal of the control is to reach the origin $(0,0)$, while minimizing the cost functional $\mathcal{J}=\int_0^{t_f}x^2(t)dt$, in which the control time $t_f$ is not fixed.

The existence of optimal solutions to the problem above is not obvious, and can be proved by showing that the Hamilton--Jacobi--Bellman equation satisfied by the value function has a classical solution (see \cite[Theorem~5.1.1 and Example~5.1.2]{schaettler-book}).

The optimal trajectories have the discrete symmetry
\[(x(t),y(t),u(t))\mapsto (-x(t),-y(t),-u({t})),
\]
and a scaling symmetry defined by the family of transformations
$$
(x(t),y(t),u(t))\mapsto (x_\lambda(t)=\lambda^2x({t}/{\lambda}),y_\lambda(t)=\lambda y({t}/{\lambda}),u_\lambda(t)=u({t}/{\lambda})),
$$
where $\lambda$ is a positive parameter. This means that if $(x(t),y(t),u(t))$ is an optimal solution then $(x_\lambda(t),y_\lambda(t),u_\lambda(t))$ is also a solution, with initial state $(\lambda^2 x_0,\lambda y_0)$ and  cost $\mathcal{J}_\lambda = \lambda^5 \mathcal{J}$.
We deduce that if  $(\bar x,\bar y)$ is a
switching point for
an optimal trajectory, i.e., the corresponding control goes from $\pm1$ to $\mp 1$ when the trajectory crosses $(\bar x,\bar y)$, then the optimal synthesis has a switching curve of equation $x=-\xi\textrm{sign}[y]y^2$, where $\xi$ is a positive constant such that the curve passes through $(\bar x,\bar y)$.

The second step of the analysis consists in applying the Pontryagin Maximum Principle, which is a necessary condition for optimality~\cite{pontryaginbook}.
The Pontryagin Hamiltonian can be expressed as
$$
H_P=p_xy+p_yu+p^0x^2,
$$
 where $(p_x,p_y)$ is the adjoint state and $p^0$  is a constant multiplier equal either to 0 or to $-\frac12$.
  If $(x(t),y(t))$ is an optimal trajectory with corresponding control $u(t)$ then
there exist  $(p_x(t),p_y(t))$ and $p_0$ such that $(p_x(t),p_y(t),p^0)\neq(0,0,0)$,
 $\dot x=\frac{\partial H_P}{\partial p_x}$, $\dot y=\frac{\partial H_P}{\partial p_y}$, $\dot p_x=-\frac{\partial H_P}{\partial x}$, $\dot p_y=-\frac{\partial H_P}{\partial y}$, and $$H_P(x(t),y(t),p_x(t),p_y(t),u(t),p^0)=\max_{v\in [-1,1]}H_P(x(t),y(t),p_x(t),p_y(t),v,p^0)=0.$$

\begin{figure}
\begin{center}
\includegraphics[scale=0.75]{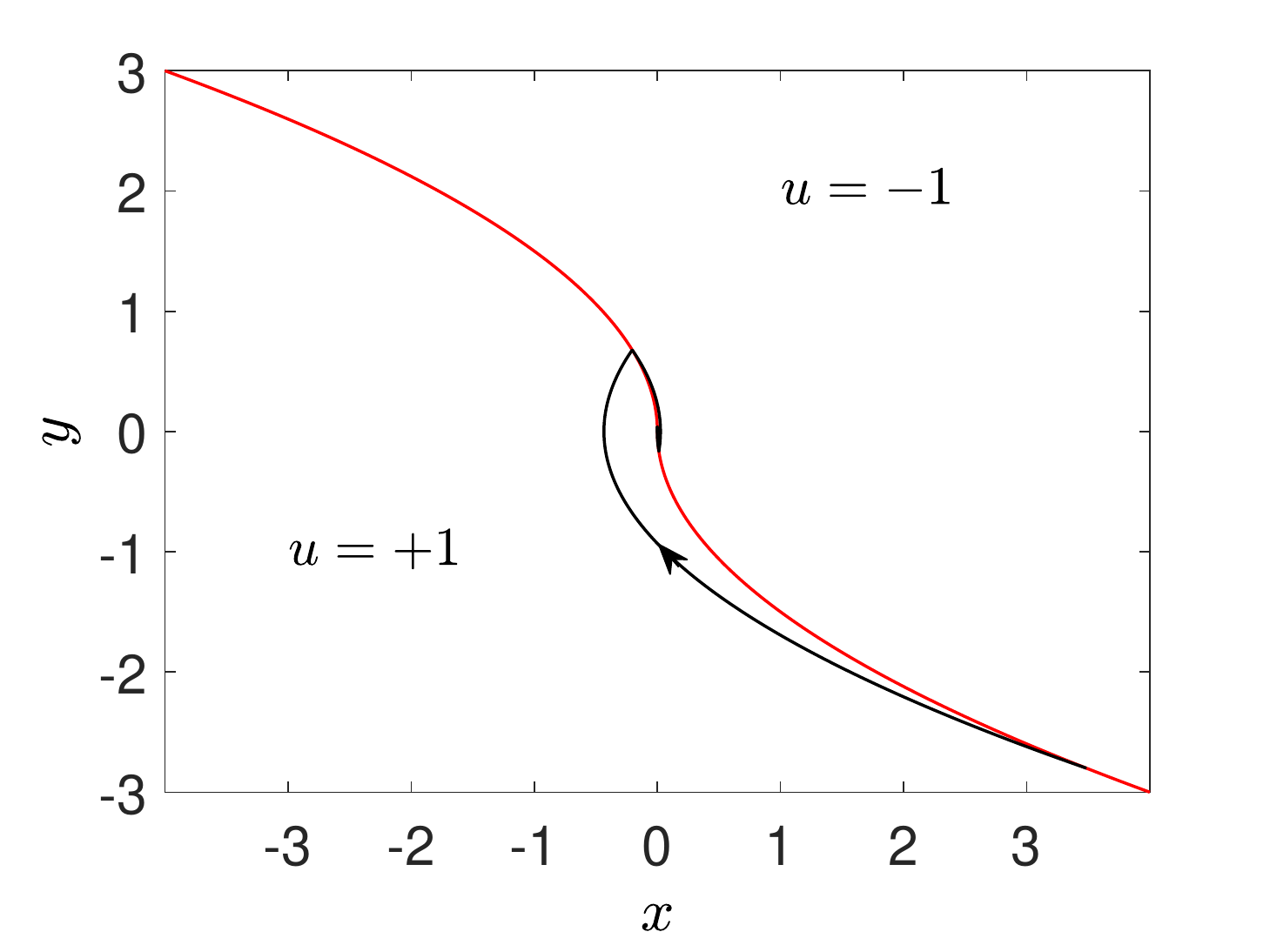}
\includegraphics[scale=0.75]{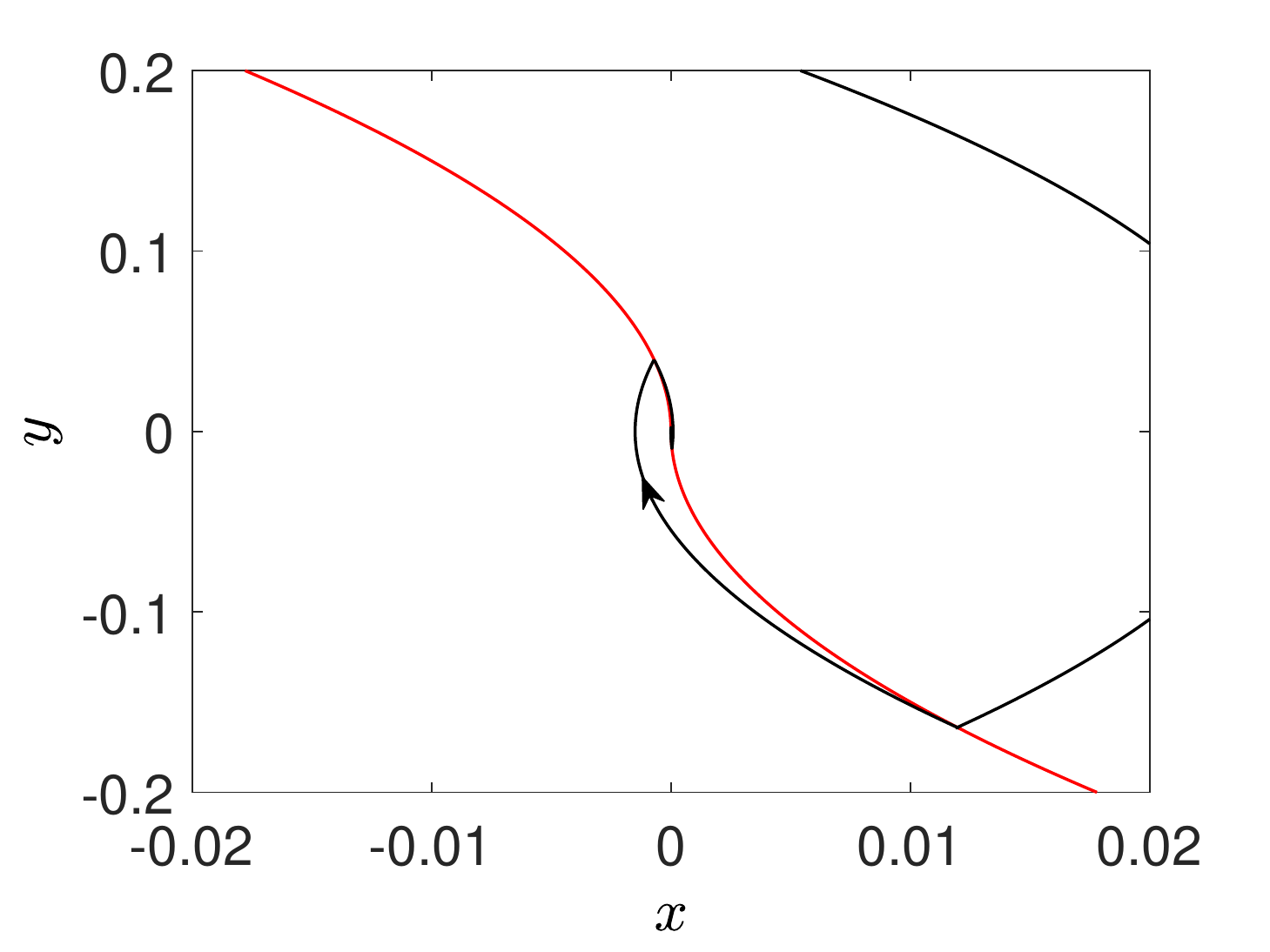}
\end{center}
\caption{(top) Optimal trajectory for the Fuller model (black solid line). The switching curves are plotted in red. (bottom) Zoom of the top panel near the origin\label{figS1}}
\end{figure}

 The switching function $\Phi$ is given by $\Phi(t)=p_y(t)$.
As in the case studied in Section~\ref{sec:PMP}, one can restrict the analysis to normal extremals.
Integrating  system (\ref{eqfuller}) from the state $(-\xi y_0^2,y_0)$ with $y_0>0$ and $u=+1$,
and imposing that the corresponding switching function vanishes at a  point $(\xi y_1^2,y_1)$ with $y_1>0$,
we obtain that $\xi$ is solution of a polynomial equation of order 4, given by
\begin{equation}\label{eq:xi}
\xi=\sqrt{\frac{\sqrt{33}-1}{24}}\simeq 0.44623.
\end{equation}
The optimal trajectory has the following symmetries:
denoting by $t_k$ the $k$-th switching time, by  $t_f^{\textrm{Ful}}$ the minimum time at which the optimal trajectory reaches $(0,0)$,
and introducing the parameter $\alpha=\sqrt{\frac{1+2\xi}{1-2\xi}}\simeq 4.13016$, we have
$$
t_f^{\textrm{Ful}}
-t_{k+1}=\frac{t_f^{\textrm{Ful}}-t_k}{\alpha},\quad \frac{x(t_{k+1})}{x(t_{k})}=-\frac{1}{\alpha^2},\quad \frac{y(t_{k+1})}{y(t_{k})}=-\frac{1}{\alpha}.
$$
Finally, starting from the state $(-\xi y_0^2,y_0)$, $y_0>0$, it can be shown that $t_f^{\textrm{Ful}}
=\frac{1+\alpha}{\alpha-1}y_0$. An example of optimal trajectory is plotted in Fig.~\ref{figS1}.

\begin{figure}
\begin{center}
\includegraphics[scale=0.75]{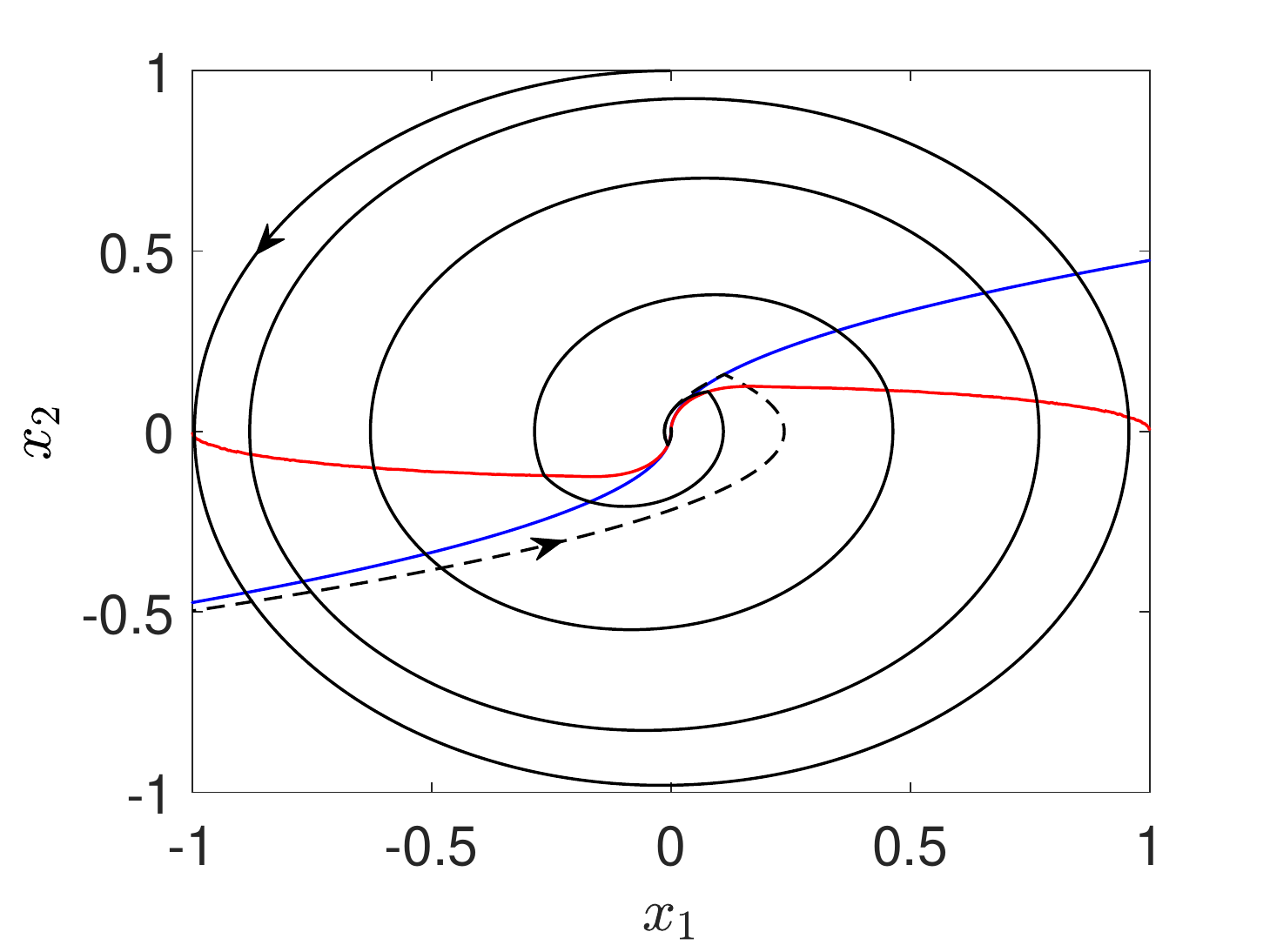}
\includegraphics[scale=0.75]{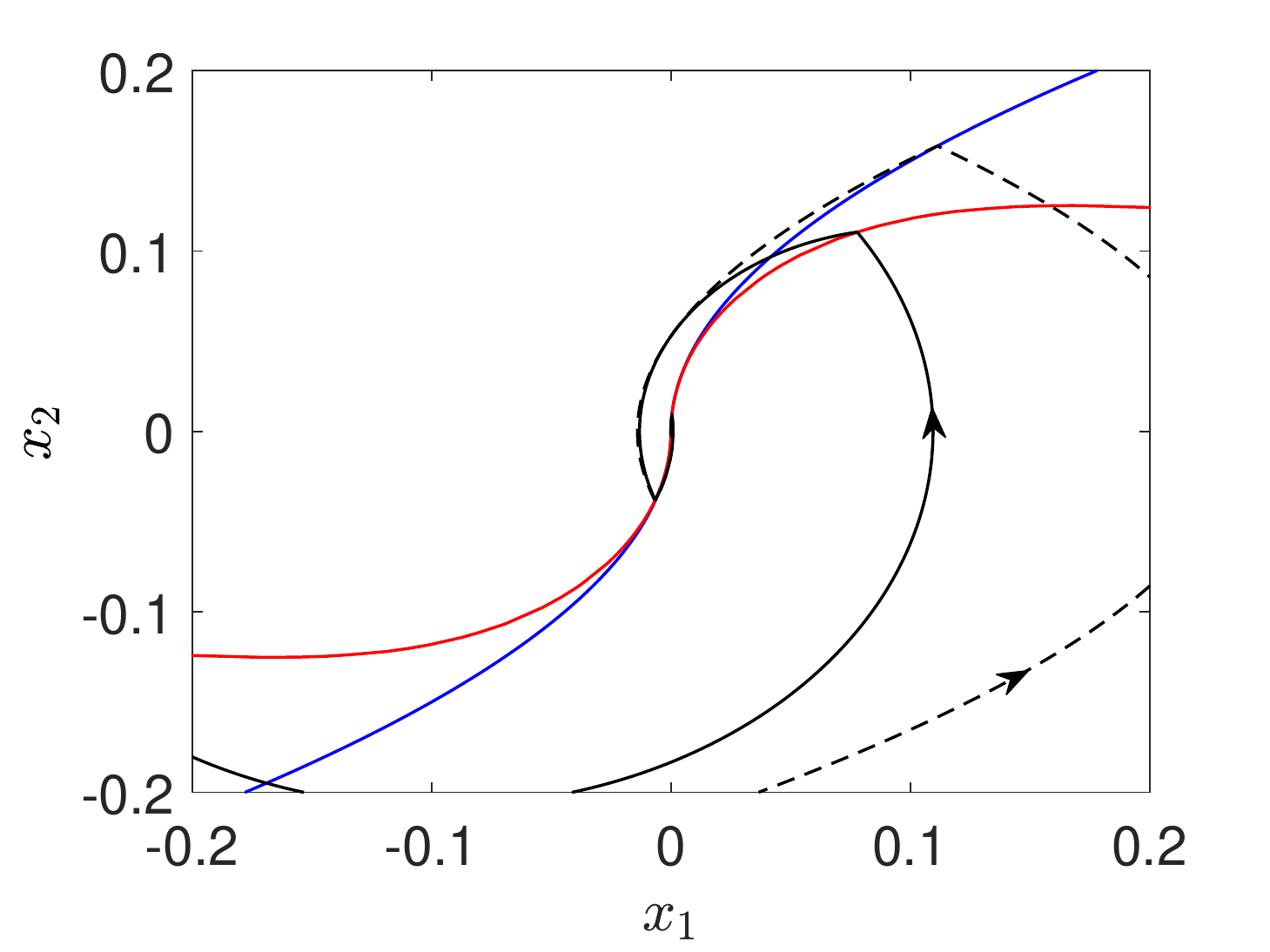}
\end{center}
\caption{Plot onto the $(x_1,x_2)$-plane of the optimal trajectory of the quantum control problem (solid black line). The dashed line depicts the solution of the approximation (4) given in the main text. The red and blue curves represent respectively the switching curves for the quantum and Fuller systems. The bottom panel is a zoom of the top one.\label{figS2}}
\end{figure}

Figure~\ref{figS2} compares the results obtained for the quantum system and its 
approximation (Eq.~(4) in the main text), which coincides with the Fuller model  up to a linear rescaling of the first coordinate. In Fig.~\ref{figS2}, we project the optimal trajectory of the quantum system onto the $(x_1,x_2)$- plane. This comparison highlights that the two solutions are very close to each other near the target state.

\section{Properties of the switching function for the three-level quantum system}\label{secswitching}
We focus in this paragraph on the switching function $\Phi$  for the three-level quantum system.
Using the Hamiltonian equations for the adjoint state in the normal case (Eq.~(2) in the main text),
one computes that $\Phi$ and its time derivatives can be expressed as
\begin{eqnarray*}
& &\Phi=p_3x_2-p_2x_3, \\
& &\dot \Phi=\Delta (x_1p_3-x_3p_1), \\
& &\ddot\Phi=-\Delta^2 \Phi+\Delta u(x_1p_2-x_2p_1)-\Delta x_1x_3.
\end{eqnarray*}
Moreover, on a segment where $\Phi\ne 0$ and, therefore,  $u$ is constantly equal to $+1$ or $-1$, one has
\begin{eqnarray*}
& & \Phi^{(3)}=-(\Delta^2+1) \dot\Phi-2\Delta u x_1x_2+\Delta^2 x_2x_3,\\
& &\Phi^{(4)}={}-(\Delta^2+1) \ddot\Phi+\Delta(\Delta^2+2)x_1 x_3
+\Delta^2 u(3x_2^2-2x_1^2-x_3^2).
\end{eqnarray*}

Note that, if $x_2\ne 0$ and $\Phi=0$, then $p_3 =\frac{x_3}{x_2}p_2$ and hence
\[\dot \Phi=\Delta \frac{x_3}{x_2}(x_1p_2-x_2 p_1).\]
Notice also that, since $H_P=0$ and  using again that $\Phi=0$, we have
\[x_1p_2-x_2 p_1=\frac{x_1^2}2\ge 0.\]

Hence, for  $x_2x_3\ne 0$ and $x_1\ne 0$, the derivative  $\dot \Phi$ has the same sign as $x_2 x_3$, which means that switches only occur  from $-1$ to $+1$ if $x_2 x_3>0$ and  from $+1$ to $-1$ if $x_2 x_3<0$. If $x_2\ne 0$ and $x_1=0$ then $\dot\Phi=0$, which implies that $p_1=0$. Therefore
$\ddot \Phi=0$ and
\[\Phi^{(3)}=\Delta^2 x_2x_3.\]
It follows that everywhere on the set $\{ x\mid x_2x_3>0\}$ switches only occur  from $-1$ to $+1$, while on $\{ x\mid x_2x_3<0\}$ they only occur  from $+1$ to $-1$. This result proves a statement of the main text, i.e. if the switching curve goes out of the north hemisphere then it passes through the points $(\pm 1,0,0)$. This property does not depend on the parameter $\Delta$.

\section{A sufficient condition for chattering}\label{seczelikin}

In this section we present an adaptation of the results presented in Chapter~3 of the book \cite{zelikinbook} by Zelikin and Borisov, concerning sufficient conditions for the appearance of Fuller-like chattering in a two-dimensional optimal synthesis.


We consider a control system of the form
\begin{equation}
    \label{eq:dynamic_sys}
    \left\{
        \begin{array}{l}
            \dot x=  \Delta y +\phi^x_1(x,y)+u\phi^x_2(x,y),\\
            \dot y= u +\phi^y_1(x,y)+u\phi^y_2(x,y).
        \end{array}
    \right.
\end{equation}
We also suppose that $\phi^x_i$, $\phi^y_i$ are smooth and small in the following  sense: denoting by $g_\kappa$  the anisotropic dilatation $g_\kappa(x,y)=(\kappa^2 x, \kappa y)$, one has
\begin{equation}\label{eq:smallness_dyn}
    \limsup_{\kappa \to 0^+} \frac{|\phi^x_i(g_\kappa(x,y))|}{\kappa} <\infty,  \qquad
    \limsup_{\kappa \to 0^+} \frac{|\phi^y_i(g_\kappa(x,y))|}{\kappa^2} <\infty,
\end{equation}
for every $(x,y)\in \mathbb{R}^2$.
Then, the  optimal control problem
\begin{equation}
    \left\{
    \begin{array}{l}
        \int_0^T x^2(t) d t \longrightarrow \min \\
        T>0\mbox{ free}, \quad u \in L^\infty([0,T],[-1,1]
        ) \\
        t\mapsto (x(t),y(t)) \in W^{1,\infty}([0,T],\mathbb{R}^2) \textrm{ solution of (\ref{eq:dynamic_sys})}\\
        (x(0),y(0))=(x_0,y_0)
    \end{array}
    \right.
    \label{pb:zelikin}
\end{equation}
satisfies the following properties.
\begin{theorem}\label{th:zelikin_2D}
For every $(x_0,y_0)$     in a sufficiently small neighbourhood of $(0,0)$, Problem (\ref{pb:zelikin}) admits a solution. Such a solution
reaches the fixed point~$(0,0)$ in finite time and its corresponding control
has infinitely many discontinuities accumulating at the final time.
Moreover, there is no other solution of  (\ref{pb:zelikin})
up to prolongation by a constant trajectory at $(0,0)$.
The optimal synthesis is characterized by a switching curve of the form
    \begin{equation}
        \Gamma=\left\{
            \begin{array}{ll}
                x=\lambda_1(y)y^2, &\quad  y>0,\\
                x=\lambda_0(y)y^2, &\quad y<0,
            \end{array}
        \right.
    \end{equation}
    where $\lambda_0$ and $\lambda_1$ are $\mathcal{C}^1$ function satisfying $\lambda_0(0)=
    -\lambda_1(0)=\xi \Delta$, where $\xi$ is as in (\ref{eq:xi}).
        The optimal control is  $-1$
    above $\Gamma$ and  $+1$
    below it.
\end{theorem}

All the proof material is in the book~\cite{zelikinbook} by M. I. Zelikin and V. F. Borisov.
Nevertheless, since this theorem is not explicitly
stated as such, we propose to retrace the ingredients of the proof and point out the relevant statements in~\cite{zelikinbook}.
\begin{proof}
    First, let us highlight the fact that the existence of a solution of~(\ref{pb:zelikin}) is not obvious. Indeed,
 it is a priori possible that
     the cost decreases by reaching the target
     later in time.
          As a consequence we are lacking compacteness for existence of the solution of the problem with free final time.
Existence of optimal trajectories can be obtained once an extremal synthesis has been constructed, using a field-of-extremals argument (\cite[Theorem~3.3]{zelikinbook}).

    The Hamiltonian given by the PMP for Problem~(\ref{pb:zelikin}) is
    \begin{eqnarray*}
       & &  H_P(x,y,p_x,p_y,p_0,u)=p_x \left(  \Delta y+\phi^x_1(x,y)+u\phi^x_2(x,y) \right)\\
       & & +p_y \left( u+\phi^x_1(x,y)+u\phi^y_2(x,y) \right)+p_0 x^2.
    \end{eqnarray*}
    Hence, the adjoint equations for normal extremals ($p_0=-1/2$) are
    \begin{equation}
        \label{eq:dynamic}
        \left\{
            \begin{array}{l}
                \dot p_x= x -p_x \partial_x \left(\phi^x_1+ u \phi^x_2 \right) -p_y \partial_x \left(\phi^y_1+u \phi^y_2 \right),\\
                \dot p_y= - \Delta p_x  -p_x \partial_y \left(\phi^x_1+ u \phi^x_2 \right) -p_y \partial_y \left(\phi^y_1+u \phi^y_2 \right) .
            \end{array}
        \right.
    \end{equation}
    If we define $(z_1,z_2,z_3,z_4)=((p_y+p_x\phi^x_2+p_y\phi^y_2)/ \Delta^2,-p_x /\Delta,-x/ \Delta,-y)$,
    we get
    \begin{equation}
        \label{eq:dynamic_z}
        \left\{
            \begin{array}{ll}
                \dot z_1 = z_2 +f_1(z,u),&
                \dot z_2 = z_3 +f_2(z,u),\\
                \dot z_3 = z_4 +f_3(z,u),&
                \dot z_4 = u +f_4(z,u),\\
                u= \, \mathrm{sign} [z_1] .
            \end{array}
        \right.
    \end{equation}
Introducing $\tilde g_\kappa(z_1,z_2,z_3,z_4)=(\kappa^4 z_1, \kappa^3 z_2, \kappa^2 z_3,\kappa z_4)$, we have the  smallness property
\begin{equation}\label{smallness-Z}
    \limsup_{\kappa \to 0} \frac{|f_i(g_\kappa(z),u)|}{\kappa^{5-i}} < \infty  \quad \textrm{ for } 1\leq i \leq 4,
\end{equation}
which generalises (\ref{eq:smallness_dyn}).
As a consequence, this system is of the form given in Equation~(3.5) of \cite{zelikinbook}.
The smallness condition (\ref{smallness-Z}) allows to apply \cite{zelikinbook}[Proposition~4.1], which guarantees, under a rank condition discussed below, that the conclusions of Theorem~3.3 in \cite{zelikinbook} hold true.

The study of this dynamical system around $z=0$ is performed in Chapter~3 of \cite{zelikinbook} thanks to a blow-up procedure. The (degenerate) hyperbolicity of the fixed point $z=0$ is established, as well as the existence of a two-dimensional invariant contracting manifold $\Sigma$ corresponding to the trajectories converging to $z=0$ {\cite[Section 3.4 - 3.8]{zelikinbook}}. This manifold is given by the trajectories of the non-smooth system (\ref{eq:dynamic_z}) switching on curves of the form
\begin{eqnarray}
& &  \hat \Gamma^0= \{(0,\mu_0(\kappa) \kappa^3, \lambda_0(\kappa)\kappa^2,\kappa)\mid \kappa<0\},\label{eq:param_hat_kappa}\\
& &   \hat \Gamma^1= \{(0,\mu_1(\kappa) \kappa^3, \lambda_1(\kappa)\kappa^2,\kappa)\mid \kappa>0\},\label{eq:param_hat_kappa2}
\end{eqnarray}
where $\mu_i,\lambda_i$ are smooth functions satisfying {(cf. \cite[Lemma 3.3]{zelikinbook})}
\begin{equation}
    \label{eq:info:lambda}
          \lambda_0(0)\in \left(-\frac{\Delta}{2},0\right),  \lambda_1(0)\in \left(0,\frac{\Delta}{2}\right),    \mu_0(0)=\frac{1}{2} \lambda_0(0)^2,   \mu_1(0)=\frac{1}{2} \lambda_1(0)^2.
\end{equation}
Besides, since $\lambda_0(0)$ and $\lambda_1(0)$ are solutions of the same polynomial system (cf.~Eq.~(3.13) in \cite{zelikinbook}) as in the case of the following Fuller dynamics
\begin{eqnarray}
    \label{eqfuller2}
   & &        \dot{x}=\Delta y, \\
   & &     \dot{y}=u,\nonumber
   \end{eqnarray}
we get $\lambda_0(0)=
-\lambda_1(0)=\xi \Delta$.

Let us now check that the projection $\pi:(z_1,z_2,z_3,z_4)\mapsto (z_3,z_4)$ restricted to $\Sigma$
 is a $\mathcal{C}^1$ mapping with a Jacobian matrix of maximal rank on $\Sigma \setminus \hat \Gamma$, as required in \cite[Theorem 3.3]{zelikinbook}. Let us denote by $\mathcal{F}(\kappa,t)$ the solution of  System~(\ref{eq:dynamic_z}) at time $t$ with initial condition parameterized by $\kappa$ as in (\ref{eq:param_hat_kappa}) (an analogous argument applying for the points of $\Sigma$ parameterized by the trajectories of System~(\ref{eq:dynamic_z}) starting from $\hat \Gamma^1$).
 Let $t$ be such that $\mathcal{F}(\kappa,t) \not \in \hat \Gamma^0\cup \hat \Gamma^1$ for all $0<s< t$.
 Then $\mathcal{F}(\kappa,s)$ is given by the solution of (\ref{eq:dynamic_z}) with $u= -1$
 and
 $\pi(\mathcal{F}(\kappa,s))$ is the solution of System~(\ref{eq:dynamic}) with constant control $u=-1$ and initial condition $(x_0,y_0)=(\lambda_0(\kappa)\kappa^2, \kappa)$.
 Therefore, $(\kappa,s)\mapsto \pi(\mathcal{F}(\kappa,s))$ is smooth. Let us check that $\partial_\kappa \pi(\mathcal{F}(\kappa,t))$ and $\partial_t \pi(\mathcal{F}(\kappa,t))$ are linearly independent. Using the fact that the flow of (\ref{eq:dynamic}) is a diffeomorphism, it is enough to check that the two vectors
\begin{equation*}
    \partial_\kappa \pi(\mathcal{F}(\kappa,0))=  (\lambda_0'(\kappa)\kappa^2+2\lambda_0(\kappa)\kappa, 1)
\end{equation*}
and
\begin{equation*}
    \partial_t \pi(\mathcal{F}(\kappa,0))
    =(\kappa + f_3(z, -1),
-1+f_4(z,-1)),
\end{equation*}
are linearly independent. Using (\ref{eq:info:lambda}), this is the case for $\kappa$ small enough. We can thus apply \cite{zelikinbook}[Theorem 3.3], which allows to conclude the proof of Theorem~\ref{th:zelikin_2D}.
\end{proof}

\section{Numerical optimization procedure}\label{sec:gradient}
\subsection{Backward stability of the optimal synthesis}
For the quantum system, we point out that the optimal trajectory can be computed numerically starting from that of the Fuller model. When we are sufficiently close to $|3\rangle$, we approximate the switching curve of the quantum system by that of its approximation. As described in the main text, we then integrate backwards in time the dynamics of the state and of the adjoint state. We stress that it makes sense to integrate backwards the extremal flow since, according to the results in~\cite{zelikinbook}, the map $\Upsilon$ associating with a switching point the subsequent one is hyperbolic and has $\Gamma$ as stable manifold towards $|3\rangle$. As a result, the $k$-th  iteration of the inverse of $\Upsilon$ tends towards $\Gamma$ as $k$ grows.
\subsection{A direct optimization method}
We apply in this paragraph a direct optimization method to design numerically the solution of the optimal control problem. We use the open access optimal solver BOCOP~\cite{bocop}. A direct optimization approach is a procedure in which the state and the time are discretized in time, transforming the initial optimal control problem into a nonlinear constrained optimization problem. In this optimization protocol, the optimal control problem is slightly modified. The goal is to minimize the cost $\mathcal{C}=\int_0^{t_f}x_1^2(t)dt$, while reaching a state as close as possible of the target $|3\rangle$ in a fixed time $t_f$ (note that the algorithm does not converge if the time is free). In order to have the fairest comparison possible between the two approaches, we estimate as follows the time of the optimal solution presented in the main text.
We assume that the control time is the sum of
the time $\tau$ taken by the quantum system to go from $(0,1,0)$ to a point
with  first two components equal to $(\xi\Delta x_{20}^2,x_{20})$, $x_{20}>0$ small, and the time
$t_f^{\textrm{Ful}}$ corresponding to the initial condition $(\xi\Delta x_{20}^2,x_{20})$ for the Fuller problem.
For $x_{20}=6.9 \times 10^{-4}$
we obtain
$\tau=2.5890$ and $t_f^{\textrm{Ful}}=0.0011$, which leads to $t_f=2.5901$. We set $t_f$ to 2.59 in the numerical optimization procedure. The time subdivision is regular and given by the number of time steps $N$, going from $N=50$ to $N=400$. The other optimization parameters are set to their default or recommended values in BOCOP.

\begin{figure}
\begin{center}
\includegraphics[scale=0.75]{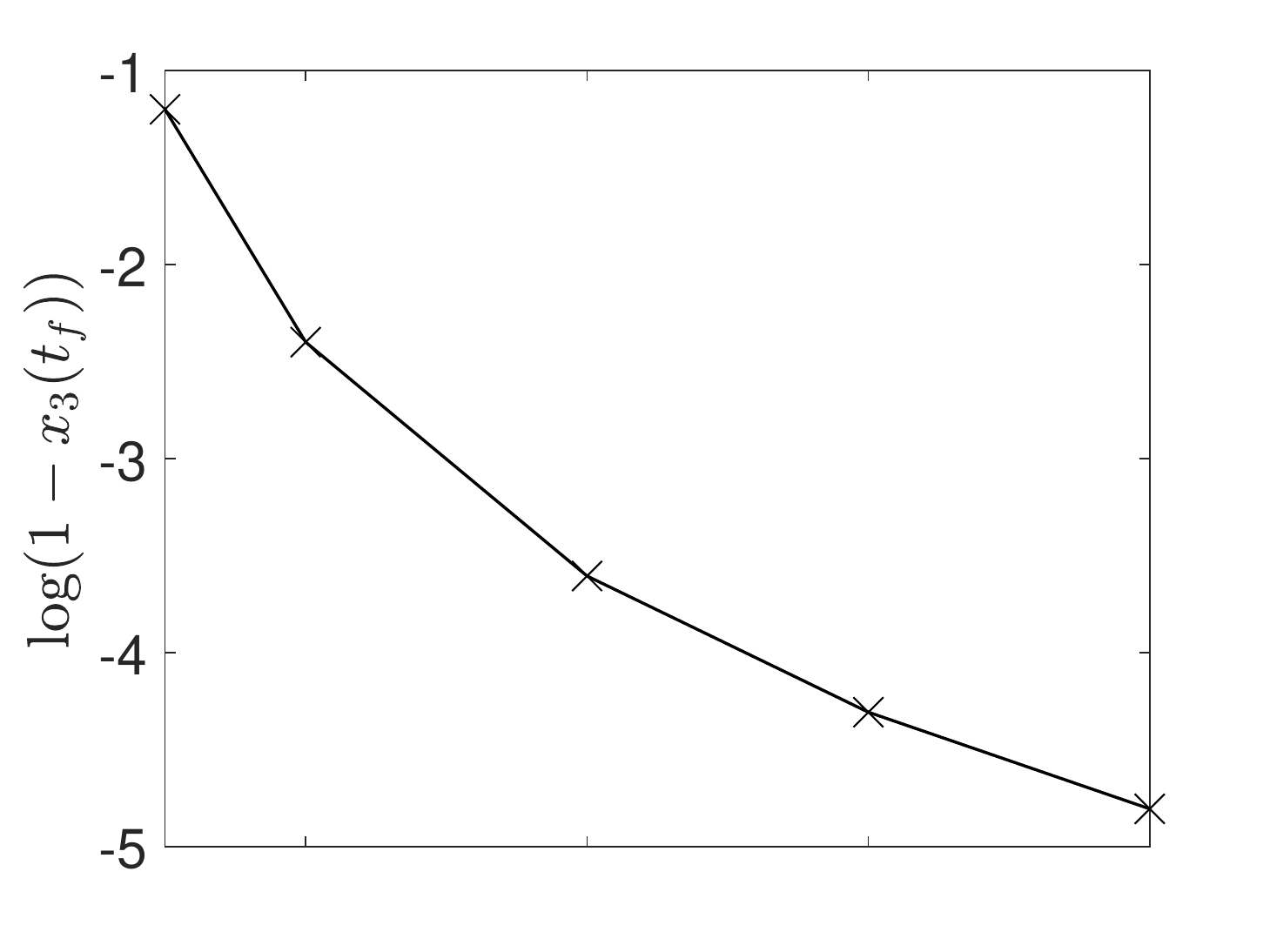}
\includegraphics[scale=0.75]{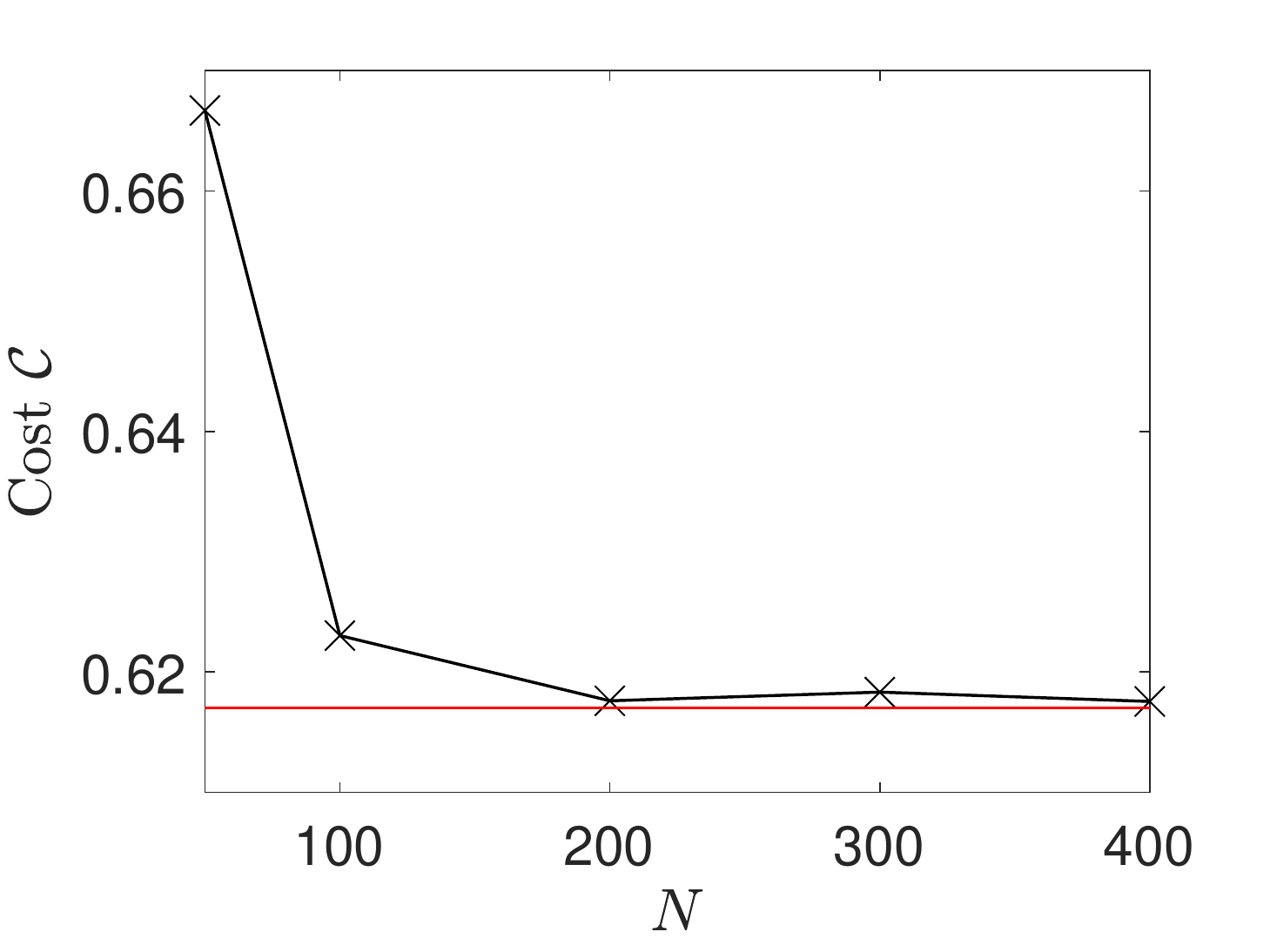}
\end{center}
\caption{Evolution of the distance to the target (top) and of the cost $\mathcal{C}$ (bottom) as a function of the number of time steps (crosses). In the bottom panel, the horizontal solid line (in red) indicates the efficiency of the optimal solution. The solid black line has been added in the two panels to  help the visualisation.\label{figS3}}
\end{figure}

In addition of the controls displayed in Fig.~5 of the main text, we report in Fig.~\ref{figS3} additional numerical results about the distance to the target and the cost $\mathcal{C}$ with respect to $N$. When $N$ becomes larger, the numerical optimal process seems to converge towards the optimal solution both in terms of final state and cost. For quite small values of $N$, we observe that the efficiency of the control is already reasonable. We point out that such sub-optimal strategy could be a possible option to bypass the chattering phenomenon in the optimization procedure.\\ \\

\end{document}